\documentstyle[preprint,aps]{revtex}
\begin{document}
\draft
\title{Are there $\eta$-Helium bound states?}
\author{S. Wycech\thanks{e-mail "wycech@fuw.edu.pl"}}
\address{Soltan Institute for Nuclear Studies,
Warsaw, Poland}
\author{A.M. Green\thanks{e-mail "green@phcu.helsinki.fi"}}
\address{ Research Institute for Theoretical Physics, P.O. Box 9,
FIN--00014 University of Helsinki, Finland}
\author{J.A. Niskanen\thanks{e-mail "janiskanen@phcu.helsinki.fi"}}
\address{Department of Physics, P.O. Box 9, FIN--00014 University
of Helsinki, Finland}

\maketitle
\begin{abstract}
Using multiple scattering theory the scattering
lengths of $\eta$ mesons on helium nuclei are calculated
and checked against final state $\eta$ interactions from the $pd
\rightarrow \eta ^3$He and $dd \rightarrow \eta ^4$He reactions. The
existence of an $\eta^4$He quasibound state is indicated.
\end{abstract}

\newpage
\section{Introduction}
\label{intro}

  In this paper we concentrate on the few-body interactions of
$\eta$ mesons. These could complement our knowledge on the
$\eta$-nucleon interaction and tell us possible evidence of
$\eta$-nuclear quasibound states. Such quasibound states were
predicted by Haider and Liu \cite{HL} and detailed calculations
performed by Li et al. \cite{Li}, when it turned out that the
$\eta$-nucleon interaction was attractive. To be observable these
states should be narrow enough, and this is not likely to happen for
the lowest $\eta$ states in large nuclei. On the other hand it was
suggested by Wilkin \cite{Wilkin} that the  rapid slope seen in the
$pd\rightarrow  \eta^3$He amplitude of Ref. \cite{GAR} just above
the threshold may signal that a quasibound  state is generated
already for small nuclei($A=3$). In contrast, a recent study of the
$dd\rightarrow \eta ^4$He  reaction shows no structure  due to any
final state $\eta ^4$He interaction \cite{Fra}. All this could
indicate a large $\eta ^3$He scattering length and a small one for
$\eta ^4$He. However, quite an opposite interpretation is put
forward in this paper. We calculate the $\eta ^3$He and $\eta ^4$He
scattering lengths and find that the former is {\it smaller} than
the latter, and that they also differ in the sign of the real part.
This suggests that the $\eta$-nucleus attraction is not strong
enough to give any binding effect in the $\eta ^3$He system,
but it is likely to give one in the $\eta ^4$He system.

   In the standard theory of final state interactions the energy
dependence of reactions is assumed to be determined by the
scattering amplitude between the final state particles
\cite{watson}. In this paper we show that the shape of the low
energy $\eta$ production cross section is also significantly
influenced by an interference of the free and the scattered waves in
the final $\eta$-helium states, because the corresponding scattering
lengths are not very large. This interference is such that the
decrease with energy becomes
steeper for both $^3$He and $^4$He than that
calculated from the final state scattering amplitude alone. However,
in the scattering amplitude itself the real and imaginary parts of
the scattering amplitude, due to the above mentioned difference in
the sign of the real parts, could be expected to  conspire so that
the slope in the $^4$He case would be somewhat smaller than for
$^3$He. Numerical results do not support this for $\eta N$
scattering lengths considered realistic.

  Before this physical interpretation of final state interactions is
discussed in Section \ref{sec3}, a formalism is developed in Section
\ref{sec2} to calculate the $\eta$-helium scattering lengths. By
some formal manipulations the multiple scattering series is summed.
The procedure used  is shown to converge quickly in the case of
$\eta$He optical potentials, which may be solved  exactly using the
Schr\"odinger equation. Then, necessary corrections to the optical
potential limit may be easily implemented by modifying the
equivalent multiple scattering series.

   The conclusions are not fully quantified since the $\eta$-nucleon
input is not determined uniquely. Also the $\eta$ production
mechanism is not under full control. Here a method for calculating
only the final state interaction  is given. However, this method is
presented in sufficient detail that a more complete comparison with
the data and the determination of the input uncertainties can be
made when measurements proposed at proton storage rings, such as the
one at Celsius, are performed. Comparing specifically with the
$\eta^3$He data we also perform an extensive variation of the
$\eta N$ scattering length in search for a constraint on it,
complementing elementary photo- or electroproduction.

\section{Scattering lengths}
\label{sec2}

       The $\eta$-helium scattering lengths are calculated in this
section. At first we consider the simplest optical model expressed
in terms of the $\eta$-nucleon scattering length. This problem may
be solved numerically, but in order to improve the method an
equivalent alternative for the optical model is provided, which
consists of a partial summation of the multiple scattering series
generated by the optical potential. The sum is expressed in terms of
multiple integrals.

      Next, some necessary improvements to the optical model are
introduced into the partial sum. These are essentially twofold:
(i) removing multiple collisions on the same nucleon and,
(ii) introducing an off-shell $\eta$-nucleon scattering matrix.
Other effects, e.g. the Pauli principle, are less significant.
These improvements introduce massive changes to the $\eta$-helium
scattering lengths defined here as the zero energy limit of
the effective range expansion
\begin{equation}
p \cot \delta = \frac{1}{a} + \frac{1}{2} \, r_0 \, p^2 \, .
\label{efran}
\end{equation}
In particular, for the $\eta ^3$He system, the
simplest optical model yields a large negative real part of the
scattering length indicating  the existence of a quasibound state.
However, when the corrections are included, a sizable positive
scattering length emerges. On the other hand, in the $\eta ^4$He
system we do find indications for  a quasibound state close to
threshold.

\subsection{Multiple scattering expansion for the inverse
scattering length}

In Ref. \cite{SW} a multiple scattering scheme was proposed to
calculate the energy shifts and widths in the atomic states of
antiprotons interacting with a light nucleus. In this
paper we apply the same method to $\eta$He scattering at
threshold. First the procedure is presented in some detail,
since it is important to understand to what extent the basic
form of the multiple scattering scheme is, in this case,
numerically equivalent to the standard optical model
approximation.

The scattering matrix $T(\eta A)$ for an $\eta$ meson
interacting with a nucleus of $A$ nucleons may be expressed
as a series  in the following way by first considering the
scattering from two non-overlapping fixed centres.
In that limit  it can be shown that the scattering matrix
at zero energy has the {\it exact} form \cite{walecka}
\begin{equation}
\label{wal}
T=\frac{T_1+T_2}{1-\frac{1}{2}\frac{(T_1+T_2)D(T_1+T_2)}{(T_1+T_2)}},
\end{equation}
where $D=1/l$ is the  propagator of the scattered particle -- with
$l$ being the distance apart of the two scattering centres  -- and
the $T_i$ are the scattering matrices from the separate centres.
This expression has the following feature that is important in few
body systems. By expanding the denominator, a multiple scattering
series emerges which -- through the factor of 1/2 in the denominator
-- automatically takes into account the exclusion of successive
scatterings from the same centre.   Being guided by this and
denoting $T=\sum T_i$, one might then naively expect the analogous
expression for the scattering from $A$ non-fixed centres to have a
form
\begin{equation}
\label{series}
T(\eta A)=<T> [1+P]^{-1},
\end{equation}
where
\begin{equation}
P=-\frac{(A-1)<TDT>}{A<T>} ,
\label{porig}
\end{equation}
with
\begin{equation}
 D=-m/(2\pi|{\bf r}-{\bf r'}|)
\label{4}
\end{equation}
the zero-energy propagator for a free $\eta$. Here the factor
$(A-1)/A$ is the generalisation of the exclusion factor 1/2
in Eq. (\ref{wal}) and $m$ is the reduced mass of the $\eta A$
system.

However, as it now stands Eq. (\ref{series}) cannot be
correct beyond $O(T^2)$, since it does not give the required form
\[T(\eta A)=<T>+\left(\frac{A-1}{A}\right)<TDT>+
\left(\frac{A-1}{A}\right)^2<TDTDT>\]
\begin{equation}
\label{mss}
+\left(\frac{A-1}{A}\right)^3<TDTDTDT>+..,
\end{equation}
 when expanded in powers of $T$. One way of
ensuring that  this correct expansion results is to modify  Eq.
(\ref{series}) to
\begin{equation}
\label{series1}
T(\eta A)=<T> [1+P+Q+R+..]^{-1},
\end{equation}
where the quantities  $P,Q,R,.. $ are of order $T, T^2, T^3,...$,
respectively,
and are chosen in turn to  guarantee Eq. (\ref{mss}).
For example, on expanding the denominator of Eq. (\ref{series1}) the
term of $O(T^2)$ is $-Q+P^2$, which in Eq. (\ref{mss}) should
give $\left(\frac{A-1}{A}\right)^2<TDTDT>$. Since $P$ has
already been fixed by the second term in Eq. (\ref{mss}), we get
$Q=P^2-\left(\frac{A-1}{A}\right)^2<TDTDT>$. This is a unique
procedure and,  neglecting for the moment the above exclusion of
consecutive scatterings on the same nucleon, it leads to the
expressions
\[
P=-\frac{<TDT>}{<T>} \ \ \ , \ \
Q=\frac{<TDT>^2}{<T>^2}-\frac{<TDTDT>}{<T>}
\]
\begin{equation}
\label{R}
R=-\frac{<TDT>^3}{<T>^3}+2 \frac{<TDTDT><TDT>}{<T>^2}
-\frac{<TDTDTDT>}{<T>} .
\end{equation}
for the first three terms of the series in the denominator. Also in
$Q,R...$ it is immediately seen that the integers multiplying each
term cancel each other e.g. in $Q$ we see +1 and --1, in $R$ we see
--1,+2,--1 etc. In the fixed centre limit all of the integrals
reduce to the same value, so that $Q,R... $ are then all zero as
found in Ref. \cite{walecka}. The important point is that
numerically this cancellation continues to a great extent even away
from the fixed centre limit as seen below. Therefore, one could hope
that the introduction of the ratios $<TDT>/<T>$ etc. of double and
triple scatterings and other "disconnected" terms at various places
would speed up the convergence of the multiple scattering series.

Here now $T$ denotes a scattering matrix of the $\eta$ from $A$
nucleons in the impulse approximation. At "zero" energy
$m T(\eta A)/2\pi$ reduces
to minus the $\eta A$ scattering length $a(\eta A)$, and
\begin{equation}
\label{3}
T=\frac{2\pi}{\mu}\, t \, A\;\rho(r)
\end{equation}
with $t$ being the $\eta$-nucleon scattering matrix at the
appropriate energy, $A\rho(r)$ the nuclear density  and $\mu $
the reduced mass for the $\eta N$ system.
The expectation values appearing in Eqs. (\ref{R}) can be expressed
in terms of the propagator and nuclear density as
\begin{equation}
\label{5}
<T>=\frac{2\pi}{\mu}tA\int d{\bf r}\,\rho({\bf r})=\frac{2\pi}
{\mu}\, tA
\end{equation}
\begin{equation}
\label{6}
<TDT>=-\frac{2\pi}{m}\,\left(\frac{m}{\mu} t\right)^2
A^2\int\int d{\bf r}d{\bf r'}\,\rho({\bf r})\frac{1}{|{\bf r}-
{\bf r'}|}\rho({\bf r'})
\end{equation}
\begin{equation}
\label{7}
<TDTDT>=\frac{2\pi}{m}\,\left(\frac{m}{\mu}t\right)^3A^3
\int\int\int d{\bf r}d{\bf r'}d{\bf r''}\,
\frac{\rho({\bf r})\rho
({\bf r'})\rho({\bf r''})}{|{\bf r}-{\bf r'}||{\bf r'}-{\bf r''}|}
,\, {\rm etc}.
\end{equation}

Using the Gaussian density profile
\begin{equation}
\rho(r)=1/(\sqrt\pi R_0)^3 \exp[-(r/R_0)^2]
\label{gauss}
\end{equation}
 one obtains now the expansion coefficients
\[
P= t\left(\frac{Am}{R_{\rm RMS} \mu }\right)\sqrt{\frac{3}{\pi}}
 \ \ , \ \ \
Q= t^2\left(\frac{Am}{R_{\rm RMS} \mu }\right)^2\left[\frac{3}{\pi}-1
\right] \ ,
\]
\begin{equation}
\label{R1}
R= t^3\left(\frac{Am}{R_{\rm RMS} \mu }\right)^3\left[
\left(\frac{3}{\pi}\right)^{3/2}
-2\left(\frac{3}{\pi}\right)^{1/2}
 +0.7796\, \sqrt{2}\left(\frac{3}{\pi}\right)^{3/2}\right],
\end{equation}
where $R_{\rm RMS}=\sqrt{3/2} R_0$ is the RMS matter radius of
the A nucleons. The number 0.7796 in $R$ is the result of a double
summation and is expected to have an accuracy of $\pm 0.0001$.
When, the terms of the series in the denominator are clustered
into increasing powers of $t$ as indicated in Eqs.
(\ref{series1},\ref{R}), it is found
that there exists a considerable amount of cancellation, e.g.
$P/d=0.977  $, $Q/d^2$=--0.045 and $R/d^3$=0.0076 where
$d=t\left(Am/R_{\rm RMS} \mu \right)$.  Therefore, if $t/R_{\rm RMS}$
is reasonably small -- as it is in the present case of
$\eta N$ scattering -- the series appears to converge rapidly.

A check on the convergence is given in Table \ref{tabopt},
where a comparison is made between the above series expansion
for $T(\eta A)$ and its value calculated directly from the
equivalent optical model potential
\begin{equation}
\label{optpot}
V({\rm opt})=\frac{2\pi}{\mu}t\cdot A\;\rho(r) \ .
\end{equation}
In this comparison the factors $(A-1)/A$ in Eq. (\ref{mss})
must be neglected. For completeness, the impulse approximation
(IA) result
\begin{equation}
\label{IA}
a(\eta A,\, {\rm IA})=-A\,\frac{m}{\mu}\, t
\end{equation}
is also quoted in Table \ref{tabopt}
[i.e. $T(\eta A)$ with $ P=Q=R=0$]. As a first
approximation $t$ is taken to be minus the $\eta N$ scattering
length, i.e. $t(E=0)=-a(\eta N)$. The actual numbers used are
a representative sample from the following sources.
Several groups  have performed coupled
channel analyses of $\eta$-nucleon and $\pi$-nucleon scattering
\cite{HL,Li,Wilkin,TIA,BL,Krusche}. These differ in the input data
and also in some details of the extraction of the $a(\eta N)$. The
$\eta N$ scattering lengths
obtained are: 0.27+i0.22 and 0.28+i0.19 \cite{BL} and
$(0.50 \pm 0.20)+i(0.33 \pm 0.06)$ fm \cite{Wilkin}. The recent
electroproduction data yield 0.476+i0.279 fm \cite{TIA}, while
photoproduction experiments suggest the possibilities: 0.430+i0.394,
0.579+i0.399, 0.291+i0.360 fm \cite{Krusche} .

The multiple scattering effect is dramatic as compared with the
impulse approximation alone, changing the attractive real part
of the $\eta N$ amplitude into a repulsive real part for the
$\eta A$ amplitude. However, the near equality between
$a(\eta A)$ and  $a(\rm opt)$ gives confidence that, indeed,
the series in Eq. (\ref{series1}) is rapidly convergent and the
use of only the terms $P,\, Q$ and $R$ gives a sufficient accuracy.

Having shown in some detail that, indeed, the standard optical
model approach can be replaced by the multiple scattering series
of Eq. (\ref{series1}), it now seems justified to modify the
latter to include effects not so easily incorporated directly
into the optical model.

\subsection{Corrections to the optical model}

Several improvements can be made to the optical model approach
and these can be
implemented into the partial sum of Eq. (\ref{series1}).

i) Firstly, as shown in Eq. (\ref{mss}) the factors
$A^{1,2,3}$ in the rescattering quantities $P,Q$ and $R$
calculated
in Eqs. (\ref{R1}) should be replaced by $(A-1)^{1,2,3}$ to
prevent the $\eta$ from interacting successively with the same
nucleon,  i.e. $a(\eta A)\rightarrow a_{A-1}(\eta A)$. In Table
\ref{tabam1} this effect is demonstrated for the $\eta N$ scattering
lengths used in Table \ref{tabopt}, and it is seen to have a large
effect in all cases. In particular, this correction makes the
real parts of the scattering lengths small for $^3$He making the
existence of a quasibound state in the $\eta^3$He  system
indicated by the optical model questionable. The absence of such
a state seems to be further confirmed by later corrections
for $^3$He. However, the real parts tend to become even more
negative in the case of $^4$He.

ii) Another improvement to the above series
is to use an $\eta N$ scattering
amplitude that is more appropriate for scattering on a bound nucleon
in a medium. This can be approximately taken into account by
extrapolating $a(\eta N)$ off the energy shell through replacing the
above scattering amplitude $a(0)$ at zero energy by $a({\rm off})$
at a negative energy defined via the equation
\begin{equation}
\label{off}
\frac{1}{a(0)}\rightarrow \frac{1}{a({\rm off})}=
\frac{1}{a(0)} -iK_{\eta},
\end{equation}
where $K_{\eta}=i\sqrt{2\mu (E_{\rm sep}+E_{\rm rec})}$ with $E_{\rm
sep,rec}$ being the $A\rightarrow (A-1)+1$ separation energy and the
recoil energy of the $\eta N$ pair relative to the residual nucleus.
For $^3$He [$^4$He] these quantities have the values $E_{\rm
sep}=7\, [21]$ MeV and $E_{\rm rec}=12\, [12]$ MeV. The effects are
shown in Table \ref{taboff}. There it is seen that our best estimate
$a_{A-1}(\eta A,{\rm off})$ of the $\eta A$ scattering length is
quite different from that predicted by the optical model. For
example in $\eta ^3$He scattering, the negative Re$\, a$(opt)
has turned positive and comparable to that of the original Impulse
Approximation. This indicates that there is no binding in
this system. However, for $\eta ^4$He the negative sign of
Re$\, a$(opt) is maintained suggesting a quasibound state.
At the end of this section these
effects are interpreted in terms of poles in the scattering matrix.
One should note in this context that, since in Eq. (\ref{off})
a significant non-zero value is assigned for the $\eta$ momentum,
the next term in the effective range expansion (\ref{efran})
could become important, if $r_0$ is large. This could have now
the effect of changing the energy variation present in Eq.
(\ref{off}).

iii) The major mechanism that generates the imaginary part of
$a(\eta A)$
is the reaction $\eta A_i\rightarrow N^*(A-1)\rightarrow \pi A_f $,
where $N^*$ is the nucleon  resonance $N^*(1535)$ with a strong
coupling both to the $\eta$ and the pion. Therefore, for $\eta$
scattering on deuterium or $^4$He -- both isoscalars -- the final
nucleus $A_f$ cannot be an isoscalar. Because the spin is not
involved in this $s$-wave scattering, then for example with
the deuteron the final $NN$ state must be
the $^3P_1$ state and also the transition operator must be spatially
antisymmetric.  This opens up the interesting
possibility that pionic inelastic
channels are damped in these cases  leading to a reduction of the
in-medium value of Im$\; a(\eta N)$. However, as shown in the
Appendix this turns out to be only a very small effect
and so this correction is not included in the present calculations.

    If there exists a pole in the scattering matrix close to the
threshold, the scattering lengths may become larger than the nuclear
radius. To some extent this situation is met here,
in particular in $^4$He. In the
case of a bound state Re$\; a < 0$, while a virtual state
corresponds to Re$\; a > 0$.  The connection is unique provided the
effective range is small, which is assumed here. However,
the validity of this assumption is not clear. Another
complication arises because of the presence of decay channels
described here by
Im$\; a(\eta N)$.  Even though there is no detailed many channel
structure of the scattering matrix, let us, however, look for the
poles given by the condition  $(1 -ip a)$= 0. With our best values
$a_{A-1}(\eta A,{\rm off})$ we have a pole in the upper complex
momentum half plane, i.e. a quasibound state in the $\eta ^4$He
case. This situation is, in fact, typical for all
Re$\; a(\eta N)$ in the range 0.3 to 0.6 fm or even higher.
On the other hand, with positive Re$\;
a_{A-1}(\eta A,{\rm off})$ one meets a virtual state  in $\eta^3$He
systems.

\section{ Final State interactions}
\label{sec3}

Since there are no beams of $\eta$-mesons, the interactions of
these mesons may be seen only via final state interactions or via
the decay mechanisms of quasibound states. As seen in Fig. 1, the
$pd\rightarrow\eta^3$He production amplitude displays a rapid
fall-off away from the threshold region, which led Wilkin to
conjecture that an $\eta^3$He quasibound or resonance state exists
nearby \cite{Wilkin}. This is reflected by the approximate
proportionality of the cross section  to the final state interaction
factor \cite{watson}
\begin{equation}
\label{CWF}
|F_1|^2=\left|\frac{a(\eta A)}{1-ipa(\eta A)}\right|^2,
\end{equation}
where $a(\eta A)$ is the $\eta$-helium scattering length and $p$ is
the $\eta$ momentum. It was found by Wilkin in the optical potential
approach \cite{Wilkin}, recalculated here in Section \ref{sec2}, that
Im$\;a(\eta ^3He)$ is rather large, which gives the required slope
and indicates a singularity. However, surprisingly the recent data
on the reaction $dd\rightarrow\eta ^4$He indicate no such slope in
the cross section close to the threshold \cite{Fra}. We now analyse
these two measurements below.

First, let us note that Eq. (\ref{CWF}) provides a good description
only, if $|a(\eta A)|\gg R_{\rm RMS}$ -- a condition not well
satisfied here by the $R_{\rm RMS}$ for $^3$He.  A more general
model needs the final $s$-state wave function for the $\eta$-He
system $\psi^- (r)$. One particularly simple form of  $\psi^- (r)$
is that from a separable potential with the Yamaguchi form factors
$(1+p^2/\beta^2)$ \cite{Yam}, which gives
\begin{equation}
\label{Wave}
\psi^- (r)=\frac{\sin (pr)}{pr}
+f^*\;\frac{[\exp(-ipr)-\exp(-\beta r)]}{r}
\end{equation}
Here $f=F_1=a(\eta A)/(1-ipa(\eta A))$ is the on-shell $\eta$-helium
scattering matrix, where the $a(\eta A)$ are taken to be the
$a_{A-1}(\eta A$, off) from Section \ref{sec2} and not the $a(\eta
A)$ given by the separable potential.  Since the factor
$[\exp(-ipr)-\exp(-\beta r)]$ determines the behaviour of the
scattered wave inside the range of the interaction, it can be
interpreted as producing an off-shell effect into the reaction. A
plausible choice of $\beta=1/R_{\rm RMS}$ is taken -- but, as shown
below, the shape of the cross section is rather insensitive to the
actual value of $\beta$.

In the reaction process the $\eta$'s are produced with some
amplitude $H(r,p_i)$ that depends both on the initial projectile
momentum $(p_i)$ and on the spatial extent of the process. For
$\eta$ energies  in the range of 0--5 MeV the dependence on $p_i$
($\approx$1 GeV) is presumably small. So far there is no complete
understanding of the actual production mechanism
\cite{Wilkin,Laget}. However, for the present purposes it is
sufficient to make only some rather general qualitative statements
concerning this mechanism.  Here we simply assume a
proportionality of the production amplitude to the nuclear density
used to derive Eq. (\ref{R1}) $H(r)=\exp\left(-(r/R_0)^2\right)$
with $R_0=\lambda \sqrt{2/3}R_{\rm RMS}$ and $\lambda\approx 1$
being a natural choice. In this way, the final state interaction
factor becomes
\begin{equation}
\label{fsi}
|F_2(\lambda, \beta)|^2=|\int \bar{\psi}^-(r)\;H(r) d\bar{r}|^2.
\end{equation}

At first sight it appears that this model for incorporating final
state interactions contains two adjustable parameters $\lambda$ and
$ \beta$. However, in practice the $\beta$ dependence is weak with
even $\beta=\infty$ being not unreasonable. As said above, we
typically fix $\beta$ at $1/R_{\rm RMS}$ leaving only  the $\lambda$
dependence. Lacking an actual model for $\eta$ production, in all
cases the results are normalised to give the experimental value of
the spin-averaged quantity
\begin{equation}
\label{expt}
|f({\rm expt})|^2=\frac{p_d}{p_{\eta}}\frac{d\sigma}{d\Omega}
(pd\rightarrow \eta^3{\rm He})=0.63\pm 0.02\,\mu{\rm b/sr}
\end{equation}
at $p_{\eta}=0.246 {\rm fm}^{-1}$.

The original hope had been that, with $\lambda$ around unity, a good
fit would be obtained to the shape of the experimental data.
However, this was only so for potential IV. In that case, with
$\lambda$=0.88 and $\beta=1/R_{\rm RMS}$ there was a very shallow
minimum in the $\chi^2$ fit to $|f({\rm expt})|^2$. It should be
added that this fit did not include the lowest experimental point at
$p_{\eta}=0.051 {\rm fm}^{-1}$, since this is thought to be subject
to large systematic errors due to beam width effects, including
energy losses in the target \cite{Wilkin}. The results are shown in
Table \ref{tabfin} and Fig. 1.

This table illustrates the following points:\\
1) As seen from columns 3 and 4 the dependence on $\beta $ is weak.
Both $\beta=1/R_{RMS}$ and $\infty$ yield good fits to the data,
since fixing $\beta=1/R_{RMS}$ gives $\chi^2$/data point = 0.35,
which is increased to only 1.02 for $\beta=\infty$.\\
2) In column 5 the use of only $|F_1|^2$ as in ref.\cite{Wilkin} is
clearly inferior with its $\chi^2$/dp = 6.63.\\
3) The normalisation factors needed to fit the experimental value of
0.63 $\mu$b/sr at $p_{\eta}=0.246{\rm fm}^{-1}$ are 0.87, 0.31, 1.50
for columns 3, 4, 5, respectively. This shows that $|F_2|^2$ is 1.7
times stronger than  $|F_1|^2$ and so could account for a
significant part of the factor of 2.5 by which the model of
Ref. \cite{FW} underestimated the experimental data.

It should be added that there is a strong correlation between
$\lambda$ and $\beta$, e.g. for $\beta=2/R_{RMS}$ the minimum
$\chi^2$/dp is still 0.35 but with $\lambda$=0.97. The dependence on
the parameter $\lambda$ is also weak as it is with $\beta$.
Therefore, the main dependence may be expected to arise from the
input values of the elementary $\eta N$ scattering amplitude.

Unfortunately, the refinement in going from   $|F_1|^2$ to $|F_2|^2$
gives less benefits with the other potential options.\\
a) For potential III with $\beta=\infty$ a $\chi^2$/dp minimum of
0.58 occurs at $\lambda=0.38$ to be compared with $\chi^2$/dp=0.71
for $|F_1|^2$ i.e. little is gained by the refinement -- in both
cases a good fit being achieved to the data. Again there is a strong
correlation between $\lambda$ and $\beta$ with the above
$\chi^2$/dp=0.58  arising also for $\beta=1/R_{\rm RMS}$ and
$\lambda=0.14$.\\
b) Potential I gives already a good fit to the data using $|F_1|^2$
with $\chi^2$/dp=0.61. This cannot be matched by $|F_2|^2$, which
gives $\chi^2$/dp=16 with $\beta=1/R_{\rm RMS}$ and $\lambda =1$.
This only improves as $\beta$ increases and $\lambda$ decreases i.e.
finally back to $|F_1|^2$.\\
c) Potential II is the worst combination. Here $|F_1|^2$ gives
$\chi^2$/dp=6.5. In comparison $|F_2|^2$ using $\beta=1/R_{RMS}$ and
$\lambda =1$ gives $\chi^2$/dp=49, i.e. neither model gives a
reasonable fit to the data.  As with potential I, this only improves
as the $|F_1|^2$ limit is approached.

The corresponding results with potential IV for $^4$He are shown
in Fig. 2. There it
is seen that $|F_1|^2$ from Eq. (\ref{CWF}) gives a visually better
fit to the data and that $|F_2(\lambda=0.88,\beta=1/R_{RMS})|^2$
appears to produce too much energy dependence. However, it should
be noted that here the experimental data have large error bars and
exist only at a few energies. In the opinion of the authors, this
should not be considered a fatal problem. Clearly some reduction of
the experimental errors would be welcome to make these data more
selective.

So far the values of $a(\eta N)$ used are those suggested by
experiment. However, these differ considerably amongst themselves
with $a(\eta N)$=[0.3--0.6]+i[0.3--0.4]fm being a more reasonable
estimate (see caption of Table \ref{tabopt}). In view of this, it is
of interest to make a global variation of the input $a(\eta N)$ to
recognize the optimal regions to fit the $pd \rightarrow \eta ^3{\rm
He}$ cross section data within different model regimes. Such a
calculation was performed for $\beta = 1/R_{\rm RMS}$ and $\lambda =
1$, i.e. in a crude model where these are not varied. Fig. 3 shows
the results for $\sqrt{\chi^2/dp}$ in the complex $a(\eta N)$ plane.
In the hatched regions this parameter is smaller than unity and
other contours show the values 2,3,...,10. It can be seen that there
is a systematic change due to each correction introduced in this
work into the optical model results -- with all these additional
effects being in the same direction. There may be a common area
around $a(\eta N) = 0.4+i0.3$ fm for the optical model \cite{Wilkin}
without a Born background introduced in Eq. (\ref{Wave}) and for the
full model, but elsewhere the models are exclusive. The $^3$He data
would allow in each model a valley of minimum $\chi^2$ in different
regions for $a(\eta N)$. So it is clear that (even assuming that the
production mechanism were known) these data cannot uniquely
determine the scattering length, although they set a strong
constraint. It may be noted that similar fits could be attempted
for the $^4$He data. However, there the quoted experimental errors
are so large that as such the fit would be useless. Even so, the
energy independence of the production amplitude indicated by the
four existing data points close to the $dd \rightarrow \eta ^4$He
threshold in very suggestive. It was not possible to produce this
feature with any reasonable value of the elementary scattering
lengths allowed by the above considered models for $^3$He.
Similar energy dependences in the $^4$He case are also obtained
by Wilkin in Ref. \cite{csb}.

\section{Conclusions}

This paper is in two distinct parts. In the first, the basic
$\eta$-nucleon scattering length $a(\eta N)$ is converted into
effective $\eta-^{3,4}$He scattering lengths $a(\eta^{3,4}$He),
which, in the second part, are then used to calculate the final
state interactions in the $pd \rightarrow \eta ^3$He and $dd
\rightarrow \eta ^4$He reactions.

The step from  $a(\eta N)$ to $a(\eta^{3,4}$He) is made in two
stages using a multiple scattering expansion, the accuracy of which
was first checked  in the optical model limit -- a limit that could
be calculated directly from the Schr\"odinger equation (see Table
\ref{tabopt}). Both the first stage, in which the replacement
$A\rightarrow (A-1)$ is made, and the second stage, in which the
scattering from a nucleon that is bound is taken into account, give
large corrections that tend to go in the same direction. The overall
effect is to give $a(\eta^{3,4}$He)'s that are very different from
those expected using the pure optical model (see Tables \ref{tabam1}
and \ref{taboff}). However, it should be added that this calculation
ignores the effect of the possible presence of a  sizable effective
range in the basic $\eta N$ interaction.

When the above $a(\eta^{3,4}$He)'s are used to extract the effect of
final state interactions from the $pd \rightarrow \eta ^3$He
reaction, it is found that only one (option IV) of the $a(\eta N)$'s
proposed in the caption of Table \ref{tabopt} is able to give a good
fit to the $^3$He data -- but not the less restrictive $^4$He data
(see Figs. 1 and 2).

In an attempt to see if there exist other values of $a(\eta N)$
that can give a good fit to $^3$He and, in addition,  give a better
fit to the $^4$He data, a search was made in the region
$1.0\geq {\rm Re}\, a(\eta N) \geq -1.0$ and
$1.0\geq {\rm Im}\, a(\eta N) \geq 0.0$.
However, this did not produce any $a(\eta N)$ significantly better
than the earlier option IV. If one may disregard the $^4$He data
either as too inaccurate or arising from too complex a reaction,
it seems that the $^3$He results indicate some potential
for constraining the elementary $\eta N$ scattering length.

\acknowledgements

This paper follows the initiative of Torleif Ericson to study the
theory and phenomenology of $\eta$ physics around CELSIUS. The
authors wish to thank Colin Wilkin very much for kindly supplying
optical model results. One of us (S.W.) wishes to thank the Research
Institute for Theoretical Physics in Helsinki, where most of this
work was carried out, and also the Swedish Institute for supporting
his earlier stay in Uppsala. J.A.N. was partly supported by the
Academy of Finland.

\appendix
\section{Isospin 0 states}

The $\eta$-deuteron and $\eta$-$^4$He systems are special cases.
These are isospin 0 systems. Therefore, some decay modes to the
pion-nucleon channels are not allowed by isospin conservation. In
the multiple scattering expansion this blocking is due to a
cancellation of pionic waves emitted from several coherent sources.
This effect has been shown to be important in coherent
$\eta$-production processes \cite{oset}.

  Here we first summarise briefly a two channel description of
$\eta$-nucleon scattering. Then we discuss the question of blocking
the pionic channel in isospin 0 systems. We
follow the standard description \cite{HL,TIA} in terms of a
separable matrix $\hat{T}$ or $\hat{V}$ dominated by coupling to the
$N^*(1535)$ resonance. Let $\hat{V}$ be
\begin{equation}
\label{sepb}
V_{ij}=\frac{f_if_j}{E-M_0},
\end{equation}
where $M_0$ is the bare mass of the $N^*$ and the $f_i$  are
couplings to the different channels. The latter are functions of the
channel momenta $q_\eta$ and $q_\pi$. The scattering matrix $\hat{T}$
follows from the Lippman-Schwinger equation
\begin{equation}
\label{LS}
\hat{T}=\hat{V}+\hat{V}(E-H_0+i\epsilon)^{-1}\hat{T}.
\end{equation}
The separability of the interaction (\ref{sepb}) then allows for
the simple solution .
\begin{equation}
\label{LSs}
T_{ij}=\frac{f_if_j}{E-M_0-\sum_k<G_k>},
\end{equation}
where
\begin{equation}
\label{LS2}
<G_k>=\int\frac{d{\bf q}}{(2\pi )^2}N(q)\frac{f_k^2(q)}{E-E_k(q)}.
\end{equation}
The value of the total Re$\; <G>$ yields
an energy shift for the N*, while Im$\; <G>$ determines its width.
With a relativistically invariant normalisation $N(q)$  for both
the $\eta$ and $N$ in Eq. (\ref{LS2})  one obtains for the
partial width into channel $k$
\begin{equation}
\frac{\Gamma_k}{2} = -{\rm Im}\;<G_k>=\frac{q_k(E)\, M_N\,
f_k^2(q_k(E))}{4\pi E}
\end{equation}
and the $T_{\eta\eta}(0)=-a(\eta N)$. The
parameters of the coupling strengths and form factor ranges may be
fitted to $\eta$-photoproduction (electro-production) and
$\pi$-nucleon scattering data as well as to the $N^*$ decay
properties. Analyses of this sort have been performed by several
groups\cite{HL,TIA,BT}. These differ slightly in the treatment of
relativistic effects and on the (uncertain) input, and there is
significant variation in the actual predictions for the scattering
lengths  $a(\eta N)$.

   Now, let us consider $\eta$ scattering on a correlated $S=1,\,
T=0 $ pair of nucleons forming a quasideuteron state. The
intermediate states in the pionic channels have $T=1$ and so, due to
the Pauli principle, the intermediate nucleons must be antisymmetric
in space coordinates. This may reduce the available phase space and
so lead to a blocking of virtual (or real) $\eta-\pi$ transitions.
As a consequence the effective Im$\; a(\eta N)$ may be reduced in a
nuclear medium. To allow for this effect we calculate the correction
to the $<G_{\pi}>$ of Eq. (\ref{LS2}) due to this Pauli effect. An
average quantity
\begin{equation}
\label{av}
\frac{1}{4}<(f_1+f_2)G_{\pi NN} (f_1+f_2)>
\end{equation}
is calculated with an antisymmetrised free $NN$ propagator -- the
average being taken over the $NN$ ground state.  Further, in this
estimate a zero-range interaction is assumed between the meson and
nucleons, which are considered to be fixed. In this way a correction
term $<\Delta G_{\pi}>$ is obtained in the form
\begin{equation}
\label{block1}
<\Delta G_{\pi}>=\int \frac{d{\bf q_\pi}}{(2\pi)^3}\, N(q_\pi)
\frac{f^2_{\pi}(q_\pi)\Delta(q_\pi)}{E-E_{\pi}(q_\pi)}\, ,
\end{equation}
where
\begin{equation}
\label{block2}
\Delta(q)=\int d{\bf u}\phi^2_{NN}({\bf u})\left[2\sin^2
(\frac{{\bf q}\cdot{\bf u}}{2})-1\right] =-\tilde{\rho}({\bf q}).
\end{equation}
Here $\phi_{NN}$ is the initial $NN$ wavefunction  and  $\tilde{
\rho}$   is the Fourier transform of the related density. For large
systems this correction disappears, since $<2\,\sin^2(\frac{{\bf
q}\cdot{\bf u}}{2})>\rightarrow 1$.  But it could be sizable, if the
inverse $R_{\rm RMS}$ of the system is comparable to the momenta
involved. However, for Im$\;<\Delta G_{\pi}>\propto q_{\pi}
\tilde{\rho}(q_{\pi})f^2_{\pi}(q_\pi)$ with
$q_{\pi}\approx 2{\rm fm}^{-1}$    one
finds only a few per cent change of the $N^*$ width in the deuteron
and in helium. This is so small a correction -- also obtained
at high momentum, where the wave functions tend to be uncertain --
that it is reasonable to neglect its effect.

\begin{figure}
\caption{The $pd\rightarrow \eta ^3$He amplitude square
$|f({\rm expt})|^2$ defined in Eq. (\protect\ref{expt}) plotted
against the $\eta$ momentum in the c.m. system for the four
elementary $\eta N$ amplitudes I--IV given in Table
\protect\ref{tabopt}. Dashed curve: optical model; Dotted:  optical
model corrected by $A \rightarrow A-1$ but described by Eq.
(\protect\ref{CWF}); Dash-dot: off-shell effect also included in Eq.
(\protect\ref{CWF}); Solid: the full model with corrections to the
optical model and with the background term in the wave function
(\protect\ref{Wave}). The data are from Refs.
\protect\cite{Wilkin,GAR,Nefkens}.}
\end{figure}

\begin{figure}
\caption{The $dd\rightarrow\eta^4$He  amplitude squared for the
elementary $\eta N$ amplitude IV. Dots: data from
\protect\cite{Fra} (normalized as the total cross section).
Curves as in Fig. 1.}
\end{figure}

\begin{figure}
\caption{Contour diagrams of $\protect\sqrt{\chi^2/dp}$ for
different models: a) optical model, b) optical model corrected by
$A\rightarrow A-1$ but described by Eq. (\protect\ref{CWF}), c)
off-shell effect also included, d) the full model with corrections
to the optical model and with the background term in the wave
function (\protect\ref{Wave}).}
\end{figure}

\begin{table}
\caption{Comparison for $\eta$-He scattering lengths $a(\eta$He) (in
fm) from various stages of the series expansion in Eq.
(\protect\ref{series1}) with the results from direct calculation
with the corresponding optical potential \protect\cite{Wilkin1}. The
numbers are for $^3$He and those in the brackets refer to $^4$He.
 The results are illustrated with four sets of the $\eta N$ input:
 $\, a(\eta N)=0.476+i\, 0.279$ fm (I),  $0.579+i\,0.399$ fm (II),
 $0.430+i\,0.394$fm (III) and   $0.291+i\,0.360$fm (IV).
In all cases $R_{\rm RMS}=1.788\, [1.618]$ fm. \protect\\}
\begin{tabular}{c|cccc|c}
$a(\eta N)$&$a({\rm IA})$&$Q=R=0$&$R=0$&$a(\eta A)$&
$a({\rm opt})$  \\ \hline
I &1.89+i1.11&--2.00+i3.01&--1.94+i2.65
&--1.89+i2.60&--1.87+i2.59  \\
&[2.63+i1.54]&[--2.46+i1.27]&[--2.14+i1.12]
&[--2.05+i1.13]&[--2.01+i1.16]  \\ \hline
II &2.30+i1.59&--2.41+i1.94&--2.16+i1.71
&--2.08+i1.70&--2.06+i1.72  \\
&[3.20+i2.20]&[--2.24+i0.83]&[--1.91+i0.79]
&[--1.81+i0.83]&[--1.79+i0.90] \\ \hline
III &1.71+i1.57&--1.67+i2.12&--1.56+i1.93
&--1.52+i1.92&--1.51+i1.93  \\
&[2.38+i2.18]&[--2.03+i1.13]&[--1.78+i1.06]
&[--1.70+i1.08]&[--1.70+i1.12] \\ \hline
IV &1.16+i1.43&--0.93+i1.92&--0.90+i1.81
&--0.89+i1.80&--0.88+i1.80  \\
&[1.61+i1.99]&[--1.62+i1.38]&[--1.47+i1.28]
&[--1.42+i1.30]&[--1.42+i1.31] \\
\end{tabular}
\label{tabopt}
\end{table}

\begin{table}
\caption{The notation is the same as Table \protect\ref{tabopt}
except that the $A^n$ factors in $P,\, Q$ and
$R$ are now replaced by $(A-1)^n$ giving $a_{A-1}(\eta A)$
as indicated by Eq. (\protect\ref{mss}). \protect\\ }
\begin{tabular}{c|ccc|c}
$a(\eta N)$&$Q'=R'=0$&$R'=0$&$a_{A-1}(\eta A)$&$a({\rm opt})$
\\ \hline
I &0.53+i4.27&0.30+i4.19&0.28+i4.16&--1.87+i2.59 \\
&[--3.01+i2.94]&[--2.76+i2.55]&[--2.67+i2.51]&
[--2.01+i1.16] \\ \hline
II &--1.52+i4.40&--1.61+i4.06
&--1.59+i4.01&--2.06+i1.72  \\
 &[--3.03+i1.84] &[--2.67+i1.64]&[--2.56+i1.65]&
[--1.79+i0.90] \\ \hline
III &--0.53+i3.35&--0.61+i3.21
&--0.60+i3.18&--1.51+i1.93  \\
 &[--2.38+i2.23] &[--2.16+i2.02]&[--2.09+i2.02]&
[--1.70+i1.12] \\ \hline
IV &--0.13+i2.36&--0.16+i2.31
&--0.16+i2.30&--0.88+i1.80  \\
 &[--1.53+i2.25] &[--1.45+i2.10]&[--1.41+i2.09]&
[--1.42+i1.31] \\
\end{tabular}
\label{tabam1}
\end{table}

\begin{table}
\caption{The notation is the same as in Table \protect\ref{tabam1}
except that $a(\eta N)$ is now calculated off-shell using
$E_{\rm sep}+E_{\rm rec}=19 \, [33]$ MeV corresponding to
$iK_{\eta} = 0.581\, [0.766]$ fm$^{-1}$. \protect\\}
\begin{tabular}{c|ccc|c}
$a(\eta N,0)$&$a(\eta N,{\rm off})$&$R'=0$&$a_{A-1}
(\eta A,{\rm off})$&$a_{A-1}(\eta A,0)$ \\ \hline
I&0.39+i0.17&2.01+i2.85
&1.99+i2.86&0.28+i4.16 \\
&[0.37+i0.15]&[--1.56+i5.30]&[--1.59+i5.19]&
[--2.67+i2.51] \\ \hline
II&0.47+i0.22&1.36+i4.38
&1.32+i4.37&--1.59+i4.01 \\
&[0.44+i0.18]&[--3.14+i3.88]&[--3.07+i3.77]&
[--2.56+i1.65] \\ \hline
III&0.39+i0.24&0.93+i3.08
&0.92+i3.07&--0.60+i3.18 \\
&[0.37+i0.21]&[--1.76+i3.69]&[--1.73+i3.62]&
[--2.09+i2.02] \\ \hline
IV&0.29+i0.26&0.59+i2.17
&0.58+i2.17&--0.16+i2.30 \\
&[0.29+i0.23]&[--0.78+i2.95]&[--0.78+i2.93]&
[--1.41+i2.09] \\
\end{tabular}
\label{taboff}
\end{table}

\begin{table}
\caption{The final state interaction factors
$|F_i(\lambda,\beta)|^2$
in units of $\mu$b/sr for the elementary
amplitude IV ($a(\eta N) = 0.291 + i0.394$), with $\lambda=0.88$
and $\beta=1/R_{\rm RMS} \ {\rm or} \  \infty$. \protect\\}
\begin{tabular}{ccccc}
$p_{\eta}$& $|f({\rm expt})|^2$ & $|F_2|^2$&
$|F_2(\beta=\infty)|^2$&$|F_1|^2$ \\ \hline
0.051&  0.53(0.02)& 1.37&1.32&1.21\\
0.115&  1.07(0.03)&1.06&1.02&0.95\\
0.166&  0.86(0.015)&0.86&0.84&0.80\\
0.202&  0.74(0.014)&0.75&0.74&0.72\\
0.246&  0.63(0.020)&0.63&0.63&0.63\\
0.295&  0.50(0.016)&0.52&0.53&0.55\\
0.337&  0.45(0.018)&0.44&0.46&0.49\\
\end{tabular}
\label{tabfin}
\end{table}


\begin{references}
\bibitem{HL}
Q. Haider and L.C. Liu, Phys. Lett. {\bf B172}, 257 (1986);
Phys. Rev. C {\bf 34}, 1845 (1986).
\bibitem{Li}
G.L. Li, W.K. Cheung and T.T.S. Kuo, Phys. Lett. {\bf B195}, 515
(1987).
\bibitem{Wilkin}
C. Wilkin, Phys. Rev. C {\bf 47}, R938 (1993).
\bibitem{GAR}
M. Garcon {\it et al.}, Proceedings of the Workshop on Spin
and Symmetry in the Standard Model, Lake Louise, Alberta,
1992, Ed.B.A. Campbell (World Scientific)
\bibitem{Fra}
R. Frascaria {\it et al.}, Phys. Rev. C {\bf 50}, R537 (1994).
\bibitem{watson} K.M. Watson, Phys. Rev. {\bf 88}, 1163 (1952).
\bibitem{SW}
S. Wycech and A.M. Green, Zeit. Phys. {\bf A344}, 117 (1992).
\bibitem{walecka} L.L. Foldy and J.D. Walecka,Ann. Phys. (NY)
{\bf 54}, 447 (1969).
\bibitem{TIA}
L. Tiator, Proc. II TAPS Workshop, Guadamar 1993, Eds. J. Diaz
and Y. Schutz (World Scientific) -- Mainz preprint MKPH-T-93-16;
L. Tiator, C. Bennhold, and S. Kamalov, Nucl. Phys. {\bf A580},
455 (1994).
\bibitem{BL}
R.S. Bhalerao and L.C. Liu, Phys. Rev. Lett. {\bf 54},
 865 (1985).
\bibitem{Krusche}
B. Krusche, Proc. II TAPS Workshop, Guadamar, 1993, Eds. J. Diaz
and Y. Schutz (World Scientific).
\bibitem{Wilkin1}
C. Wilkin, private communication.
\bibitem{Nefkens}
B. Nefkens, private communication;
R. Kessler, Ph.D. thesis UCLA 1992.
\bibitem{FW}
G. F\"aldt and C. Wilkin, "The reaction $pd\rightarrow
\eta^3$He near threshold", Uppsala University preprint
TSL/ISV--94--105.
\bibitem{Laget}
J.M. Laget and J.F. Lecolley, Phys. Rev. Lett.{\bf 61},
 2069 (1988).
\bibitem{BT}
 C. Bennhold and H. Tanabe, Nucl. Phys. {\bf A530}, 625 (1991).
\bibitem{oset}
B. Lopez Alvaredo and E. Oset, Phys. Lett. {\bf B324}, 125 (1994).
\bibitem{Yam}
Y.Yamaguchi, Phys.Rev. {\bf 95}, 1628 (1954).
\bibitem{csb}
C. Wilkin, Phys. Lett. {\bf B331}, 276 (1994).
\end{references}
\end{document}